\documentclass[%
aps,%
prl,%
twocolumn,%
superscriptaddress,%
showpacs,%
preprintnumbers,%
byrevtex%
]{revtex4}

\usepackage{color,amssymb,graphicx}

\renewcommand{\vec}[1]{{\bf #1}}
\renewcommand{\paragraph}[1]{\textcolor{red}{{{\small {\bf {\sc#1}}}}} \\ }

\def\ee{\mathrm{e}}
\def\ii{\textrm{i}\,\!}
\newcommand{\vt} {\vartheta}
\newcommand{\bea} {\begin{eqnarray}}
\newcommand{\eea} {\end{eqnarray}}
\newcommand{\beann} {\begin{eqnarray*}}
\newcommand{\eeann} {\end{eqnarray*}}
\newcommand{\labs} {\left\vert}
\newcommand{\rabs} {\right\vert}

\newcommand{\lrb} {\left(}
\newcommand{\rrb} {\right)}

\newcommand{\lab} {\left\langle}
\newcommand{\rab} {\right\rangle}

\begin{document}
\date{November 9, 2001}

\preprint{\texttt{mpi-pks/0111001}}
\title{Backbone-induced semiconducting behavior in short DNA wires}

\author {Gianaurelio Cuniberti}
\email[E-mail: ]{cunibert@mpipks-dresden.mpg.de}
\author{Luis Craco}
\affiliation{Max Planck Institute for Physics of Complex Systems, D-01187 Dresden, Germany}

\author{Danny Porath}
\affiliation{The Physical Chemistry Department, The Hebrew University of Jerusalem, IL-91904 Jerusalem, Israel}

\author{Cees Dekker}
\affiliation{Department of Applied Physics, Delft University of Technology,
NL-2628 CJ Delft, The Netherlands}

\begin{abstract}
We propose a
model hamiltonian for describing
charge transport through short homogeneous double stranded DNA molecules.
We show that the hybridization of the overlapping $\pi$ orbitals
in the base-pair stack coupled to the backbone
is sufficient to predict the existence of a gap in the nonequilibrium
current-voltage characteristics with a minimal number of parameters.
Our results are in a good agreement with the
recent finding of semiconducting behavior in short poly(G)-poly(C) DNA oligomers.
In particular, our model provides a correct description of the molecular resonances
which determine the quasi-linear part of the current out
of the gap region.
\end{abstract}

\maketitle

The attempt to understand the mechanism of electron motion along DNA is the source of an intense debate in the biochemical
and chemical physics communities%
~\cite{RJ02eco}.
Solving this problem
is an essential step for the development of DNA-based molecular
electronics.
New insights to this issue are brought by recent breakthroughs in
direct measurements through DNA molecules~\cite{KKGRVKB01,RAPKVLX01,RMPMS01,dePMHCGHHBOSA00,PBdeVD00,LLLMHW00eco,BESB99}.
Transport measurements through nanostructured systems are potentially capable
of addressing the basic issues of the conduction properties of molecular and supramolecular aggregates.
The aftermath for the realization of molecular electronics devices is
straightforward~\cite{LXZ99}.
It is thus not surprising that DNA molecules became the subject of 
an intense study concerning their potency
to carry an electric current~\cite{PBdeVD00,dePMHCGHHBOSA00,LLLMHW00eco,RAPKVLX01,BESB99,KKGRVKB01},
and
to provide a scaffold for the metal assembling of highly conductive nanowires%
~\cite{RMPMS01,BESB99}.
From a nanoelectronics perspective,
the DNA possesses ideal structural and molecular-recognition properties, and the understanding of the charge transport through DNA may result in the ambitious goal of self assembling nanodevices with a definite molecular architecture%
~\cite{RBCDiFMMSGG00eco}.

The hypothesis that double stranded DNA supports charge transport as a linear chain of overlapping $\pi$ orbitals located on the stacked base-pairs, already advanced in the early sixties%
~\cite{ES63eco},
received first experimental boosts only recently via long range electron transfer measurements~\cite{HHB96}.
As far as transport through DNA is concerned, the available experiments
are still controversial mainly due to the complexity of the environment and the molecule itself (sequence variability~\cite{MM-BG98}, thermal vibrations...).
Concerning theory, the most reliable procedure to tackle these systems would be the {\it ab initio} quantum chemistry approach. However, massive numerical costs complicate its use for realistic biological systems~\cite{SLH96}.
To our knowledge, at the present time, only few density-functional-theory (DFT) calculations for DNA molecules are
available%
~\cite{MOAS-PS99eco,dePMHCGHHBOSA00}.
In a parallel development
particular aspects of the DNA transport phenomenology have been
explained as 
mediated by polarons~\cite{CR00},
solitons%
~\cite{Lakhno00eco},
electrons or holes~\cite{RJ02eco,BPR97}.
Such lack of a unifying theoretical scheme calls for
reproducible and unambiguous experimental results which are still a great technological challenge.

Recently, Porath {\it et al.}~\cite{PBdeVD00} have reported nonlinear transport measurements on 10.4 nm long polyguanine-polycytosine DNA, corresponding to 30 consecutive GC base-pairs, attached to platinum leads (GC-device).
The measured room temperature current-voltage ($I$-$V$) characteristics show typical semiconducting features 
with a gap of the order of 1~V.
Furthermore, the poly(G)-poly(C) DNA molecule has typical electronic features of a periodic chain, as the first DFT calculations have indicated~\cite{dePMHCGHHBOSA00}.
This may support the idea that, differently from natural $\lambda$-DNA (complex sequence), where
the sequence variability or the attachment to the surface 
could lead to electron
localization over very few base-pairs~\cite{cohen}, in short suspended GC-devices band-like conduction
might be the relevant transport mechanism. 

Motivated by such considerations, in this Letter, we introduce a minimal model for charge transport through GC-devices
and show that the semiconducting behavior of the observed low temperature $I$-$V$ curves can be explained by the hybridization of the G-G $\pi$ stack with the transversal backbone reservoirs.
The HOMO-LUMO structure, as estimated by Refs.~\onlinecite{dePMHCGHHBOSA00,BBSR01,YLY98}, suggests that hole injection into the GC-devices might be fairly described using a tight-binding model by a hamiltonian comprising three terms
$H=H_\mathrm{mol} + H_\mathrm{leads} + H_\mathrm{coupl}$.
We describe here a short
poly(G)-poly(C) DNA molecule ($N=30$
base-pair long)~\cite{PBdeVD00}
as three-band model given by
\begin{figure}[t]
\centerline{\includegraphics[width=.99\columnwidth]{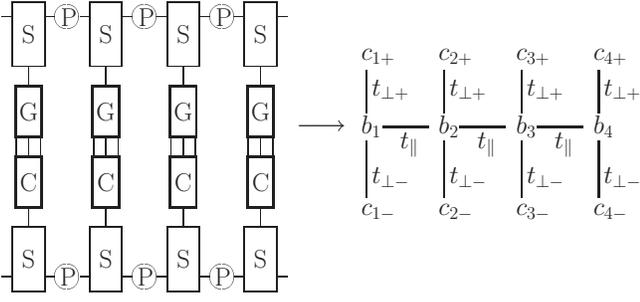}}
\caption{\label{fig:sketch} Schematic view (left) of a fragment of poly(G)-poly(C) DNA molecule; each GC base-pair is attached to sugar and phosphate groups forming the molecule backbone. On the right side, the diagram of the lattice adopted in building our model, with the $\pi$ stack connected to the isolated states denoted as $\pm$-edges.}
\end{figure}
\noindent
\beann
\label{eq:model}
H_\mathrm{mol}
&=&\varepsilon_{b} \sum_{i, \sigma} b^\dagger_{i \sigma} b^{\phantom{\dagger}}_{i \sigma}
-t_{\parallel} \sum_{\lab i, j \rab, \sigma} b^\dagger_{i \sigma}
b^{\phantom{\dagger}}_{j \sigma}
\\ &+&
\sum_{i, \sigma, \alpha=\pm}
\varepsilon_\alpha c^\dagger_{i \sigma \alpha} c^{\phantom{\dagger}}_{i \sigma \alpha}
- \sum_{i, \sigma, \alpha =\pm}
t_{\perp \alpha}
\lrb
c^\dagger_{i \sigma \alpha} b^{\phantom{\dagger}}_{i \sigma} + \mathrm{h.c.}\rrb
\nonumber
\eeann
where $b_{i \sigma}^\dagger$ is the creation operators for
charges with spin $\sigma$ in the G base site $i$ ($i,j = 1,\dots, N$),
and $c_{i \sigma \alpha}^\dagger$ the one in the $\alpha$-edge. The latter accounts for the upper sugar group sites and, possibly, for the C bases with the relative lower strand sites, as schematized in Fig.~\ref{fig:sketch}.

Charges can propagate along the $\pi$ orbital stack via the nearest neighbour hopping probability $t_\parallel$, or be hybridized to $\alpha$-edges by $t_{\perp \alpha}$;
$\varepsilon_{b}$ ($\varepsilon_{\alpha}$) is the energy level of the localized $b$ ($c_\alpha$) charge.
We have fixed the number of transversal paths per central site to two
($\alpha=\pm$). Models with more transversal hoppings are all equivalent (via a
canonical transformation). The numerical value of $t_{\perp}$ is then
consequently renormalized by the number of such transversal paths.
The leads are described by the hamiltonian
\bea
H_\mathrm{leads} = \sum_{\vec{k}, \sigma, \nu=\mathrm{L,R}} \varepsilon^{\phantom{\dagger}}_{\vec{k} \nu} a^\dagger_{\vec{k} \sigma \nu} a^{\phantom{\dagger}}_{\vec{k} \sigma \nu},
\eea
where $\vec{k}$
denotes the wave vector,
and $\varepsilon_{\vec{k} \nu}$ describes the single-electron dispersion relation of the
$\nu$-lead, measured with respect to the Fermi energy $\mu_\nu$ ($\mu_\mathrm{L,R}=\pm eV/2$).
The coupling between the leads and the molecule ends
can be described by a tunneling amplitude $U_{\vec{k}}$
of the electron in the state $(\vec{k},\sigma,\nu)$ of the $\nu$-lead to the molecule-end sites:
\bea
H_\mathrm{coupl} = - \sum_{\vec{k}, \sigma} U^{\phantom{\dagger}}_{\vec{k}} \lrb a^\dagger_{\vec{k} \sigma \mathrm{L}} b^{\phantom{\dagger}}_{\phantom{\vec{k}} \!\!\!1 \sigma} + a^\dagger_{\vec{k} \sigma \mathrm{R}} b^{\phantom{\dagger}}_{\phantom{\vec{k}} \!\!\!N \sigma}
+\mathrm{h.c.}
\rrb.
\eea

The effect of the metal pad on the molecule is given by the self-energy
\beann
\Sigma_\nu = 2 \sum_{\vec{k}} \labs U_{\vec{k}} \rabs^2 G_\nu \lrb {\vec{k}} , E \rrb ,
\eeann
where $ G_\nu \lrb {\vec{k}} , E \rrb$ is the retarded Green function for the isolated $\nu$-lead, and the factor two accounts for the spin degeneracy.

The transmission function, $T=4 \Delta_\mathrm{L} \Delta_\mathrm{R} \labs {\cal G}_{1N}\rabs^2$, is obtained by making use of the Fisher-Lee
relation~\cite{FL81}.
Here $\Delta_\nu=-\mathrm{Im}\Sigma_\nu$ is the spectral density of the metal
molecule coupling. For notational convenience, we write down the
relations for identical metal pads, $\Sigma_\mathrm{L} = \Sigma_\mathrm{R}= \Sigma$.
${\cal G}_{1N}$ is the molecular Green function between the two contact sites dressed by the lead self-energy.
The calculation of ${\cal G}_{1N}$ can be pursued analytically~\cite{%
CFR01c} leading to
\beann
\frac{\xi_{0}(\Phi)}{t_\parallel {\cal G}_{1N}}
=
{\xi_N (\Phi)}
-
2 \frac{\Sigma}{t_\parallel} {\xi_{N-1}(\Phi)}
+
\frac{\Sigma^2}{t_\parallel^2}
{\xi_{N-2}(\Phi)} ,
\eeann
with
$\xi_{N}(\Phi) =
{\lrb \Phi + \sqrt{\Phi^2 -1} \rrb^{N + 1} - \lrb \Phi -
\sqrt{\Phi^2 -1} \rrb^{N + 1}}.
$
\\
The relevant one-particle Green function at every site in the $\pi$ stack has been renormalized according to the hybridization with the backbone states so that
\bea \label{eq:gcii}
\Phi= \frac 1 {2 t_\parallel} \lrb {\cal G}^{-1}_{b} - \sum_{\alpha=\pm} t_{\perp \alpha}^2 {\cal G}_{\alpha} \rrb
\eea
where ${\cal G}_{\eta} = \lrb E + \ii 0^+ - \varepsilon_\eta \rrb^{-1}$ are the bare (isolated) Green function for the three site classes ($\eta=b,\pm$).

The backbone coupling, Eq.~(\ref{eq:gcii}), controls the opening of a gap in the transmission. This can be intuitively understood within the standard treatment for the leads.
In fact,
for
bulky electrodes, in the wide band limit, expressing the spectral density in units of $t_\parallel$, $\Sigma=-\ii \delta t_\parallel$, the transmission can be written as 
\beann
T=\frac{
4 \delta^2 \sin^2 (\vt)}{
\lrb
\sin(N+1)\vt
- \delta^2
\sin(N-1)\vt
\rrb^2
+4
\delta^2
\sin^2 N\vt
}.
\eeann
$T$ is an even function of $\Phi =: \cos \vt$ with $N$ resonances at values of the molecular orbitals broadened by the dimensionless lead spectral density $\delta$. For simplicity, we consider here equal strengths in the backbone
coupling, $t_{\perp} := t_{\perp +} = t_{\perp -}$, which does not imply any physical assumption if $\varepsilon_\eta \equiv 0$ for $\eta=b$ (preserving the charge neutrality) and for $\eta =\pm$ (no gating). In the absence of such a
backbone coupling, $\Phi$ reduces to the energy of the incoming charge
\begin{figure}[t]
\centerline{\includegraphics[width=.8550\linewidth]{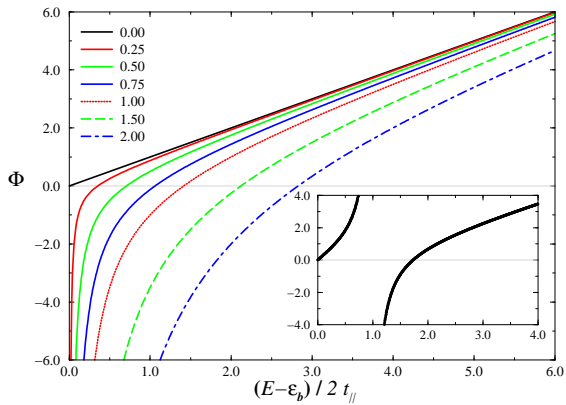}}
\caption{\label{fig:renormal} Energy renormalization due to the backbone coupling. Different lines correspond to different values of $t_\perp / 2 t_\parallel$, here $\varepsilon_+ = \varepsilon_- = \varepsilon_b$. The inset shows the case $t_\perp = 2 t_ \parallel$ where the two backbone are further gated with opposite sign $\varepsilon_\pm = \varepsilon_b \pm t_\parallel$. All curves are antisymmetric in $\Phi (t_\perp=0) = (E-\varepsilon_b) / 2 t_\parallel$.}
\end{figure}
\noindent
relative to the G base on-site energy and in units of the band width of the $\pi$
band, $\Phi (t_\perp =0) = (E-\varepsilon_b) / 2 t_\parallel$;
the transmission of a $N$-atom molecular wire is recovered~\cite{%
CFR01c}.
When a finite backbone coupling is considered,
the energy is renormalized through Eq.~(\ref{eq:gcii}), thus a gap, $\Delta_{T}$, is opened in the transmission
following the lines of Fig.~\ref{fig:renormal}.
In the absence of gating, the transmission gap reads
\bea
\label{eq:gap}
\Delta_T = 2 \sqrt{t_\parallel^2 +2 t_\perp^2} - 2 t_\parallel,
\eea 
while the width of each of the two side bands is $2 t_\parallel$, and thus
independent of $t_\perp$.
This behavior can be also understood by referring to the dispersion relation,
$\Phi (E)= - \cos q$, in the limit of an infinite wire ($N \gg 1$), $q$ being the longitudinal momentum in units of lattice spacing. The 
absence of electronic states between the two emerging branches, the HOMO-LUMO gap of our model hamiltonian, determines the gap in the transmission probability.
As to strengthen the intuition of this high reflectivity near zero energies, 
one can regard 
our model as the extreme discretization of
a phase coherent quantum waveguide with a fish bone shape (right hand side of Fig.~\ref{fig:sketch}), where
a low energy incoming particle has a high probability to be localized on the
side transepts.

\begin{figure}[t]
\centerline{\includegraphics[width=.8910\linewidth]{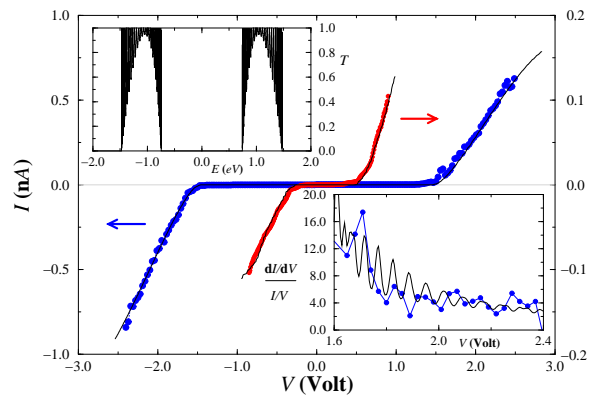}}
\caption{\label{fig:th_data} Low temperature $I$-$V$ characteristics of two typical measurements at $18$~K (blue circles) and at $3.6$~K (red circles). Solid lines show the theory curves following the experimental data.
The insets show the transmission calculated after the blue data (upper) and the
normalized differential conductance (lower).
The parameters used are $t_\parallel = 0.37$~eV and $t_{\perp \pm} = 0.74$~eV for the blue measurement, and $t_\parallel = 0.15$~eV and $t_{\perp \pm} = 0.24$~eV for the red one~\protect\cite{distributeit}.}
\end{figure}

The calculation of the current can be pursued within the scattering formalism%
~\cite{FG99eco}
\bea
I = \frac{2e}{h} \int dE\, T(E) \lrb f_\mathrm{L}(E)-f_\mathrm{R}(E) \rrb , \label{eq:curr}
\eea
where the Fermi functions $f_\nu(E)=1/( \ee^{\beta(E-\mu_\nu)}+1 )$ are controlled by the lead electrochemical potentials $\mu_\mathrm{L,R}$.
Eq.~(\ref{eq:curr}) is a reasonable estimation for the truly nonlinear current
when the bridge system is a finite molecular
chain. This has been recently shown by nonequilibrium
Green function calculations (Keldysh formalism) through onedimensional dot arrays%
~\cite{BWG01eco}.
Moreover, since we consider here
the low temperature current-voltage measurements from the experiment in Ref.~\onlinecite{PBdeVD00} (two representative
examples are plotted in Fig.~\ref{fig:th_data}),
the fit to the current can be tuned by comparing the experimental differential conductance with the transmission function (the upper inset of Fig.~\ref{fig:th_data} shows the calculated transmission for the data displayed in blue).

Theoretical results, plotted as solid lines in Fig.~\ref{fig:th_data}, show a good overall agreement with the description of the gap and the
molecular energy levels along the almost linear part of the experimental $I$-$V$ characteristics. 
Theoretical parameters have been obtained by a $\chi^2$ minimization over $t_\parallel$, constraining $t_\perp$ to give the observed experimental current gap. 
The latter is principally induced by the coupling $t_\perp$ to the side sites (see Eq.~(\ref{eq:gap})) and reproduced into a 
gap in the current-voltage curves as a result of the integration in Eq.~(\ref{eq:curr}).
For a fixed value of $t_\parallel$ the current gap is an increasing function of
$t_\perp$. That is why the smaller
hopping parameters used for the fit to the red data in Fig.~\ref{fig:th_data} correspond to the smaller gap curve, again with a
gratifying match between the theory and the experiment.
The on-site energies have been assumed all zero as to implement the presence
of counterions on the negatively charged backbone; the consequent induced dipole yet supports our fish bone construction.
As a further test for the applicability of the present model, we have analyzed the position of the molecular levels 
by comparing the theoretical and experimental normalized differential
conductance, finding a fairly good accord (lower inset in Fig.~\ref{fig:th_data}).
Note that the gap in the blue curve does not show a pronounced voltage asymmetry as observed in other experimental curves, such as the red data.  The latter measurement was performed at $3.6$~K and exhibits a smaller gap of $0.8$~eV. Here, we coped with the asymmetry by assuming different voltage drops at the molecule-electrode junctions (bias shift of $0.06$~eV) as in Ref.~\onlinecite{DTHRHK97}. 

Let us now briefly discuss other possible gap opening mechanisms. 
Electron correlation may be a source for a gap
in quasi-onedimensional systems, but for the experiment at hand
band insulator mechanisms prevail on charge Mott one~\cite{Craco99b}. Here, we have deliberately avoided the weak coupling regime since estimations of the device capacitance would lead to eventual Coulomb blockade gaps of only fractions of the observed gaps; this suggests that if the GC-device is in the strong coupling regime most of the gap is due to the molecule itself (HOMO-LUMO gap).
Other possible gap opening mechanisms have been excluded from our model because of their marginality to 
{
short} (10 nm long) or 
{
low-temperature} GC-devices. In fact
({a}) localization effects due to the sequence variability~\cite{MM-BG98},
({b}) twiston motions~\cite{BBSR01},
({c}) possible static disorder or local defects~\cite{cohen}, and
({d}) dephasing~\cite{LY01}
are all potential causes 
which may concur in determining the absence of current in some of
the recent experiments~\cite{SvanNdeVD01,dePMHCGHHBOSA00,BESB99}
performed at room-temperature on long DNA wires.

Finally, we would also like to comment on the observed gap-width variability, even within the same sample at different measurement sweeps. A structural fluctuation in the nucleoside distribution along the double helix~\cite{SRB91} may interfere with the $\pi$ stack~\cite{BGOR00} leading to a re-calibration of the overlap integrals
which indeed drives the gap-width and induces a sharp change of the $I$-$V$
profile.
Moreover, the measurement process itself may induce structural rearrangement of the double helix.
The strong electric field associated with the high nonlinear voltage drops can be responsible for a polarization of the molecule.
A possible insertion of ions may result
in different distributions of the on-site energies and the hopping integral, possibly locally, and leave a signature in the measured gap variability.
Our fits show that a change of
$t_\perp$ and/or $t_\parallel$ at one site along the chain is sufficient to induce such a current-gap
width
change or a ``switch'' in the shape of the $I$-$V$ curve~\cite{PBdeVD00}, in agreement with the
structural fluctuation hypothesis.

Further, joint theoretical and experimental work on DNA molecules would definitely contribute to a better discrimination among the possible concomitant conductance mechanisms, to check the influence of lead contacts on the device characteristics, and to qualify both the sample-to-sample gap variability and its temperature dependence.
New experiments may eventually be used to test whether low energy states can be added to the transmission.
This could be a way to check our prediction
via, e.g., changing the inter-base coupling by
doping the molecule with metal ions~\cite{RAPKVLX01,aich}.

In summary, we have considered charge transport through a short poly(G)-poly(C) DNA molecule attached to nanoelectrodes by considering the hybridization of the $\pi$ stack
with backbone states. In doing so, we have reached a 
quantitative agreement with data taken from the
experiment reported in Ref.~\onlinecite{PBdeVD00}.

\begin{acknowledgments}
LC would like to thank P. Fulde for his kind hospitality in Dresden at the beginning of 2001 when this work has been conceived.
We are indebted to Ralf Bulla, Erez Braun, Joshua Jortner, and Jos{\'e} M. Soler for fruitful discussions.
GC research at MPI is sponsored by the Schloe{\ss}mann Foundation; DP research was supported by the FIRST Foundation 
and by European grant 
IST-2000-29690.
\end{acknowledgments}

\vspace{-.6cm}

{

}

\begin{thebibliography}{10}
\bibitem{distributeit}
These values are considered for an homogeneous system. If one draws the backbone coupling parameters for a disordered system their average value would be lower. For a related experiment supporting this idea cf.\ Ref.~\onlinecite{cohen}.
\end{thebibliography}
\end{document}